
\input phyzzx
\catcode`\@=11 
\def\papers{\papersize\headline=\paperheadline\footline=\paperfootline}
\def\papersize{\hsize=40pc \vsize=53pc \hoffset=0pc \voffset=1pc
   \advance\hoffset by\HOFFSET \advance\voffset by\VOFFSET
   \pagebottomfiller=0pc
   \skip\footins=\bigskipamount \normalspace }
\catcode`\@=12 
\papers
\def\IR{{\hbox{{\rm I}\kern-.2em\hbox{\rm R}}}}
\def\IB{{\hbox{{\rm I}\kern-.2em\hbox{\rm B}}}}
\def\IN{{\hbox{{\rm I}\kern-.2em\hbox{\rm N}}}}
\def\IC{{\ \hbox{{\rm I}\kern-.6em\hbox{\bf C}}}}
\def\IP{{\hbox{{\rm I}\kern-.2em\hbox{\bf P}}}}

\def\IZ{{\hbox{{\rm Z}\kern-.4em\hbox{\rm Z}}}}

\def\d{{\rm d}}
\def\underarrow#1{\vbox{\ialign{##\crcr$\hfil\displaystyle
{#1}\hfil$\crcr\noalign{\kern1pt
\nointerlineskip}$\longrightarrow$\crcr}}}
%
\def\d{{\rm d}}
\def\ltorder{\mathrel{\raise.3ex\hbox{$<$}\mkern-14mu
             \lower0.6ex\hbox{$\sim$}}}
\def\lesssim{\mathrel{\raise.3ex\hbox{$<$}\mkern-14mu
             \lower0.6ex\hbox{$\sim$}}}

\tolerance=5000
\def\ov{\overline}

\sequentialequations
\overfullrule=0pt
\twelvepoint
\nopubblock
\line{\hfill IASSNS-HEP-93/36}
\titlepage
\title{SPACE-TIME TRANSITIONS IN STRING THEORY
\foot{Lecture at Strings, 1993 (Berkeley, May, 1993)}}

\author{Edward Witten
\foot{Research supported in part by NSF Grant PHY-92-45317.}}
\vskip.1cm
\centerline{{\it School of Natural Sciences,}}
\centerline{{\it Institute for Advanced Study,}}
\centerline{{\it Olden Lane,}}
\centerline{{\it Princeton, N.J. 08540}}
\abstract{Simple mean field methods can be used to describe transitions
between different space-time models in string theory.  These include
transitions between different Calabi-Yau manifolds, and more exotic
things such as the Calabi-Yau/Landau-Ginzberg correspondence.
}
\endpage
\REF\wit{E. Witten, ``Phases of $N=2$ Models in Two
Dimensions,'' (IASSNS-HEP-93/3), to appear in Nucl. Phys. B.}
\REF\asp{P. Aspinwall, D. Morrison, and B. Greene,
``Multiple Mirror Manifolds and Topology Change in String Theory''
(IASSNS-HEP-93/4), Phys. Lett. {\bf 303B}, 249 (1993.)}
Today I will be talking about transitions between different space-time
models in string theory.  In string theory, space-time is represented by
a two dimensional quantum field theory.  We will therefore study
transitions among such two dimensional theories that occur as the
parameters are varied.  I will describe very simple mean field methods
that can be used [\wit] to relate
Calabi-Yau models with different target spaces (as recently developed in
a different way by P. Aspinwall, D. Morrison, and B.
Greene [\asp]).  The same
methods also give a new explanation of the familiar relation between
certain Calabi-Yau models and certain Landau-Ginzburg models.

In fact, I believe that in this way we get the generalization of the
Calabi-Yau/Landau-Ginzburg correspondence to arbitrary Calabi-Yau sigma
models.  The general story involves an extension of the sigma model
moduli space beyond the classical region.  If $X$ is a Calabi-Yau
manifold, the Kahler class of a Ricci-flat Kahler metric defines a point
in $H^2(X, \IR)$.  This point lies in a conical region of $H^2(X,
\IR)$ called the Kahler cone.   In classical field theory, the region of
$H^2(X, \IR)$ outside the Kahler cone does not have much use.  In string
theory, however, (or in a world-sheet quantum field theory), nothing is
wasted.  The quantum moduli space is all of $H^2(X,\IR)$ -- and in fact,
$H^2(X,\IC)$, to allow for theta angles.  $H^2(X,\IR)$ is divided into cones,
of which the classical Kahler cone of $X$ is only one.  In each cone,
the theory has a different ``geometrical'' description.  For instance,
the traditional Calabi-Yau/Landau-Ginzburg correspondence arises when
$H^2(X,\IR)$ is one dimensional.  There are then only two cones: the
positive half-line, which is the Kahler cone of $X$; and the negative
half-line, which is the Landau-Ginzburg region.

Let me illustrate these ideas in the Calabi-Yau/Landau-Ginzburg case.
Let $z_1,\ldots,z_5$ be complex variables, and let $F(z_1,\ldots,z_5)$
be a homogeneous quanta polynomial, with the transversality property
that the equations
$$
0 = {\partial F \over \partial z_1} = \dots = {\partial F \over \partial
z_5}
\eqn\cat
$$
have a common solution only at the origin.  A generic homogeneous $F$
has that property.

Consider the hypersurface $X$ of solutions of $F=0$ in $\IC \IP^4$.  It
is a smooth Calabi-Yau manifold by virtue of the transversality of  $F$.
The sigma model with target space $X$  is an $N=2$ superconformal field
theory.

On the other hand, we can introduce chiral superfields $\Phi_1,\ldots,
\Phi_5$ in $N=2$ superspace in two dimensions, with superpotential
$$
W(\Phi_1,\dots,\Phi_5) = F(\Phi_1, \dots, \Phi_5).
\eqn\dog
$$
With this superpotential, we can make an $N=2$ model that is not
conformally invariant
$$
{\cal L} = \int \d^2\, x\, \d^4\, \theta\, \sum_i \Phi_i \Phi_i
- \left(\int \d^2\, x\, \d^2\, \theta F(\Phi_i) + h.c.\right)
\eqn\horse
$$
The ordinary potential is then
$$
V = \sum_i\left|{\partial F \over \partial \Phi^i}\right|^2 ~.
\eqn\pig
$$
This model has (by virtue of transversality of $F$) an isolated vacuum
at $\Phi_i = 0$ with massless particles; it is called a Landau-Ginzburg
model.

It is claimed (by Martinec and by Greene, Vafa and Warner) that a
suitable orbifold of this Landau-Ginzburg model is ``equivalent'' in the
infrared to the Calabi-Yau model determined by the same $F$.

To obtain a new insight about this relationship, we consider a U(1)
gauge theory in N=2 superspace, described by a vector superfield.  The
gauge invariant field strength is $\Sigma = \{ \ov{\cal D}_+, {\cal D}_-\}
/2\sqrt2$ and the gauge kinetic energy is
$$
{\cal L}_{\rm gauge} = - {1 \over 4e^2} \int \d^2x\,
\d^4\, \theta \,\ov{\Sigma} \Sigma \,.
\eqn\bear
$$
Such a $U(1)$ gauge theory in two dimensions has two additional
interactions possible, the $\theta$ angle and the Fayet-Iliopoulos $D$
term.  These can be written as
$$
L_{D,\theta} = {it \over 2\sqrt2} \, \int\d^2x\, \d\,\theta^+\,\d\,
\bar\theta^- \, \Sigma +\, h.c.
\eqn\lion
$$
with
$$
t = ir + {\theta \over 2\pi}
\eqn\tiger
$$
$\theta$ is the usual $\theta$ angle, and $r$ is the coefficient of the
$D$ term.

To this we add chiral superfields -- the five superfields $\Phi_i$ of
the earlier discussion, which we now take to have charge $1$, and a new
superfield $P$ of charge $-5$.  Their kinetic energy is
$$
{\cal L}_{kin} = \int \d^2x\,\d^4\,\theta\biggl\{ \sum_i\,\bar \Phi_i\,
\Phi_i + \bar P \, P\biggr\}\,.
\eqn\horse
$$
Finally, we introduce the superpotential, which we take to be the gauge
invariant function $W = PF(\Phi_i)$.  The corresponding piece of the
Lagrangian is hence
$$
{\cal L}_W = - \int \d^2x\,\d^2\,\theta \, P F(\Phi_i) - h.c.
\eqn\polar
$$

After performing the $\theta$ integrals and eliminating the auxiliary
fields, the ordinary potential of the model is
$$
V = {e^2 \over 2} \left(\sum \bar\varphi_i \varphi_i - 5 \bar p p -
r\right)^2 + \left|F(\varphi_i)\right|^2 + \sum_i \bar p p
\left|{\partial F \over \partial \varphi_i}\right|^2
\eqn\snow
$$
Here $\varphi_i, p$ are the bosonic components of superfields $\Phi_i,
P$.  What remains is to study the vacuum structure as a function of $r$.

For $r >> 0$, vanishing of $V_1 = {e^2 \over 2}
(\sum \bar \varphi_i\, \varphi_i - 5 \bar p p - r)^2$ requires
$\varphi_i \not= 0$. Vanishing of $V_2 = \bar p p \sum_i
\left|{\partial F \over \partial \varphi_i}\right|^2$ then implies
(given transversality of $F$) that $p = 0$.  Vanishing of $V_1$ then
requires
$$
\sum \bar \varphi_i \varphi_i = r
\eqn\sunny
$$
and after dividing by the gauge group $U(1)$, this gives a copy of $\IC
\IP^4$ with Kahler class proportional to $r$.  Finally, vanishing of
$V_3 = |F(\varphi_i)|^2$ shows that the space of classical ground states
is the Calabi-Yau hypersurface $F = 0$ in $\IC \IP^4$.

Now for $r << 0$, vanishing of $V_1$ gives $p \not= 0$.  Vanishing of
$V_2$ gives then $\varphi_i = 0$, so there is up to gauge transformation
a unique classical ground state with $p = \sqrt{-r/5}$.  In expanding
around this ground state, the $\varphi_i$ are massless and governed by
an effective superpotential  $\widetilde W = \langle p \rangle F
(\varphi_i)$.  The expectation value of $p$ breaks the gauge group to
$\IZ_5$, and the low energy theory is a $\IZ_5$ orbifold of a
Landau-Ginzburg theory with superpotential $W$.

So Calabi-Yau and Landau-Ginzburg arise as two different ``phases'' of
one system.  To probe more deeply, one must understand what happens near
$r = 0$.  For this I refer to my paper [1].  Suffice it to say that as
long as the $\theta$ angle is non-zero and at least for quantities such
as Yukawa couplings, there is a smooth continuation from $r > 0$ to $r <
0$.  The transition involves passing through a situation where stringy
effects are big and field theory is not a good approximation.

So at least to this extent, Calabi-Yau and Landau-Ginzburg are two
different limits of the same system, rather than two different phases
in the strict sense.  They are related like water to steam.  This is the
sense in which Calabi-Yau and Landau-Ginzburg are ``equivalent.''

By changing the gauge group and the quantum numbers of the chiral
superfields, one can work out many generalizations of this.  Among other
things one gets transitions among space-times of different topology.  For
more detail, I refer to [1, 2].

I find these results fascinating because along with phenomena such as
the $R \leftrightarrow 1/R$ duality, they are among relatively few
examples of really ``stringy'' phenomena that we know.  I am sure there
is much more lurking under the surface, waiting to be unearthed once we
have understood the geometrical tools and language appropriate to string
theory.  For the time being, not very much of this is accessible, though
we can hopefully do more even with the methods we already have.

The phenomenon of topology change makes me wonder about the relation of
the diffeomorphism group ${\it diff}(X)$ of a classical space-time to the
corresponding symmetry group -- call it $str(X)$ -- of string theory
with target space $X$.  (It could be that $str(X)$ is an equivalence
relation not generated by a group action, and conceivably well-defined
only on shell.  I will ignore such questions here.)  Can ${\it diff}(X)$ be
embedded in $str(X)$?  One might suppose so, but there is no evident
way to make this embedding in any formalism I know of.  I would tend to
believe that ${\it diff}(X)$ cannot be embedded
in $str(X)$; such an embedding
is certainly possible in the long distance limit, but there may be
deviations of order $\alpha'$.  The topology-changing phenomena show
that under certain conditions $str(X) = str(X')$ for distinct
space-times
$X$ and $X'$.  So if ${\it diff}(X)$ can be embedded in $str(X)$, so can be
${\it diff}(X')$, presumably, as $X$ and $X'$ seem to be on an entirely
equal footing.

One can ask the opposite question: is there a homomorphism from
$str(X)$ onto ${\it diff}(X)$?  I would presume the answer is no; the
existence of such a homomorphism would more or less give a way to
recover field theoretic physics from string theoretic physics, and any
such attempt ought to be thwarted by terms of order $\alpha'$.

At any rate, the topology-changing phenomena ought to mean that if there
is a homomorphism from $str(X)$ onto ${\it diff}(X)$, there should also be
one onto ${\it diff}(X')$.  This seems even less plausible.

The topology-changing phenomena really ought to mean that the relation
between $str(X)$ and ${\it diff}(X)$ is purely a long-distance approximation
that is relevant in a particular limit of the string theory moduli
space.  It is tantalizing to think that the deviation of the symmetry
group of the world from ${\it diff}(X)$ might have some observable effect in
the low energy world; but it is hard to see how that would be.

\refout
\end